 \documentclass[12pt,preprint]{aastex}
\slugcomment{\it Accepted by AJ}
\lefthead{Griffith et al.}

\usepackage{graphicx} 
\usepackage{amssymb}
\usepackage{amsmath}
\usepackage{anysize}
\usepackage{rotating}
\usepackage{multirow}
\usepackage{float}
\usepackage{rotating}

\def\spose#1{\hbox to 0pt{#1\hss}}
\def\simlt{\mathrel{\spose{\lower 3pt\hbox{$\mathchar"218$}}
     \raise 2.0pt\hbox{$\mathchar"13C$}}}
\def\simgt{\mathrel{\spose{\lower 3pt\hbox{$\mathchar"218$}}
     \raise 2.0pt\hbox{$\mathchar"13E$}}}

\begin{document}

\title{Spitzer Photometry of WISE-Selected Brown Dwarf and Hyper-Luminous Infrared Galaxy Candidates}

\author{R\textsc{oger} L. G\textsc{riffith}\altaffilmark{1}, 
J. D\textsc{avy} K\textsc{irkpatrick}\altaffilmark{1}, 
P\textsc{eter} R. M. E\textsc{isenhardt}\altaffilmark{2},
C\textsc{hristopher} R. G\textsc{elino}\altaffilmark{1}, 
M\textsc{ichael} C. C\textsc{ushing}\altaffilmark{3},
D\textsc{ominic} B\textsc{enford}\altaffilmark{4},
A\textsc{ndrew} B\textsc{lain}\altaffilmark{5},
C\textsc{arrie} R. B\textsc{ridge}\altaffilmark{6},
M\textsc{artin} C\textsc{ohen}\altaffilmark{7},
R\textsc{oc} M. C\textsc{utri}\altaffilmark{1},
E\textsc{milio} D\textsc{onoso}\altaffilmark{1},
T\textsc{homas} H. J\textsc{arrett}\altaffilmark{1},
C\textsc{arol} L\textsc{onsdale}\altaffilmark{8},
G\textsc{regory} M\textsc{ace}\altaffilmark{9},
A. M\textsc{ainzer}\altaffilmark{2},
K\textsc{en} M\textsc{arsh}\altaffilmark{10},
D\textsc{eborah} P\textsc{adgett}\altaffilmark{11},
S\textsc{ara} P\textsc{etty}\altaffilmark{9},
M\textsc{ichael} E. R\textsc{essler}\altaffilmark{2},
M\textsc{ichael} F. S\textsc{krutskie}\altaffilmark{12}, 
S\textsc{pencer} A. S\textsc{tanford}\altaffilmark{13},
D\textsc{aniel} S\textsc{tern}\altaffilmark{2}, 
C\textsc{hao}-W\textsc{ei} T\textsc{sai}\altaffilmark{1}, 
E\textsc{dward} L. W\textsc{right}\altaffilmark{9},
J\textsc{ingwen} W\textsc{u}\altaffilmark{2},\&
L\textsc{in} Y\textsc{an}\altaffilmark{1}}

\altaffiltext{1}{Infrared Processing and Analysis Center, California Institute of Technology, Pasadena, CA 91125}
\altaffiltext{2}{Jet Propulsion Laboratory, California Institute of Technology, 4800 Oak Grove Dr., Pasadena, CA 91109}
\altaffiltext{3}{Department of Physics and Astronomy, MS 111, University of Toledo, 2801 W. Bancroft St. Toledo, OH 43606-3328}
\altaffiltext{4}{NASA Goddard Space Flight Center, Greenbelt, MD 20771}
\altaffiltext{5}{Department of Physics and Astronomy, University of Leicester, LE1 7RH Leicester, UK}
\altaffiltext{6}{California Institute of Technology MS249-17, Pasadena, CA 91125, USA}
\altaffiltext{7}{Monterey Institute for Research in Astronomy, 200 8th Street, Marina CA 93933, USA}
\altaffiltext{8}{National Radio Astronomy Observatory, 520 Edgemont Road, Charlottesville, VA 22903}
\altaffiltext{9}{Astronomy Department, University of California Los Angeles, Los Angeles, CA 90095}
\altaffiltext{10}{School of Physics and Astronomy, Cardiff University, Cardiff CF24 3AA, UK}
\altaffiltext{11}{NASA Goddard Space Flight Center, Greenbelt, MD 20771}
\altaffiltext{12}{Department of Astronomy, University of Virginia, Charlottesville, VA, 22904}
\altaffiltext{13}{Department of Physics, University of California Davis, One Shields Ave., Davis, CA 95616}

\keywords{stars: brown dwarfs -- stars: photometry -- galaxies: photometry -- galaxies: high redshift}

\begin{abstract}

We present {\it{Spitzer}} 3.6 and 4.5 $\mu$m photometry and positions for a sample of 1510 brown dwarf candidates identified by the WISE All-Sky Survey. Of these, 166 have been spectroscopically classified as objects with spectral types M(1), L(7), T(146), and Y(12); Sixteen other objects are non-(sub)stellar in nature. The remainder are most likely distant L and T dwarfs lacking spectroscopic verification, other Y dwarf candidates still awaiting follow-up, and assorted other objects whose {\it{Spitzer}} photometry reveals them to be background sources.  We present a catalog of {\it{Spitzer}} photometry for all astrophysical sources identified in these fields and use this catalog to identify 7 fainter (4.5 $\mu$m $\sim$ 17.0 mag) brown dwarf candidates, which are possibly wide-field companions to the original WISE sources. To test this hypothesis, we use a sample of 919 {\it{Spitzer}} observations around WISE-selected high-redshift hyper-luminous infrared galaxy (HyLIRG) candidates. For this control sample we find another 6 brown dwarf candidates, suggesting that the 7 companion candidates are not physically associated. In fact, only one of these 7 {\it{Spitzer}} brown dwarf candidates has a photometric distance estimate consistent with being a companion to the WISE brown dwarf candidate. Other than this there is no evidence for any widely separated ($>$ 20 AU) ultra-cool binaries. As an adjunct to this paper, we make available a source catalog of $\sim$ 7.33 $\times 10^5$ objects detected in all of these {\it{Spitzer}} follow-up fields for use by the astronomical community. The complete catalog includes the {\it{Spitzer}} 3.6 and 4.5 $\mu$m photometry, along with positionally matched $B$ and $R$ photometry from USNO-B; $J$, $H$, and $K_s$ photometry from 2MASS; and $W1$, $W2$, $W3$, and $W4$ photometry from the WISE all-sky catalog. 

\end{abstract}

\section{Introduction}

In the historical nomenclature of stellar classification, astronomers have classified stars according to their spectra, finally settling on the following spectral types, OBAFGKM, with O 
representing the hottest stars and M the coolest \citep{morgan1943}. With the discovery of numerous (sub)stellar objects cooler than the coldest M dwarfs, the need for new spectral types became evident. To describe and characterize this new population two new spectral types were introduced to the stellar taxonomy, L and T dwarfs (see e.g. \cite{kirkpatrick2005} for an in depth review), while a new type, ``Y", is used for even colder objects, as suggested by  \cite{kirkpatrick99} and  \cite{kirkpatrick2000}.

These low-mass, low-temperature astrophysical sources are considered to be the lowest-mass population of star-like objects. Those 
with masses below $\sim$ 0.075 M$_\sun$ are referred to as ``brown dwarfs"(\citealt{kumar63}; \citealt{hayashi63}), and have core temperatures insufficient to sustain hydrogen fusion. Brown dwarfs are important objects for several reasons. They provide benchmarks with which to test our theories of gas chemistry and cloud formation in low-temperature atmospheres \citep{burgasser08}. As analogs of extra-solar planets brown dwarfs are easier to study because they are typically isolated. They represent {\it in situ} time capsules of star formation, since their mass is never recycled back into the interstellar medium, therefore preserving information on metallicity enrichment over the lifetime of the Galaxy \citep{burgasser08}. They provide important constraints on both the functional form of the stellar mass function and the low-mass limit of star formation (\citealt{kirkpatrick12}; \citealt{metchev08}; \citealt{reyle2010}; \citealt{burningham2010}). Though predicted to exist in the 1960's, brown dwarfs remained elusive for decades until surveys such as the Two Micron All-Sky Survey (2MASS; \citealt{skrutskie2006}), the Sloan Digital Sky Survey (SDSS; \citealt{york2000}), and the Deep Near-Infrared Survey of the Southern Sky (DENIS; \citealt{epchtein97}), revealed them in large numbers.

Recently researchers have shifted their focus toward the coldest brown dwarfs, which are one of the two primary science objectives of the{\it{Wide-field Infrared Survey Explorer}} (WISE; \citealt{wright2010}). WISE launched on 2009 December 14 and began surveying the sky on 2010 January 7 at wavelengths of 3.4, 4.6, 12, and 22 $\mu$m, hereafter referred to as $W1$, $W2$, $W3$, and $W4$, respectively. It completed its first full pass of the sky on 2010 July 17 and the WISE All-Sky Data Release was issued on 2012 March 14\footnote{\tt http://wise2.ipac.caltech.edu/docs/release/allsky}. The survey reaches 5$\sigma$ point source sensitivities in unconfused regions to better than 0.08, 0.11, 1 and 6 mJy, respectively \citep{wright2010}. The $W1$ and $W2$ filters were specifically designed to probe the deep 3.3 $\mu$m CH$_4$ absorption band in the spectra of brown dwarfs and the region relatively free of opacity at $\sim$ 4.6 $\mu$m, making their $W1-W2$ colors very red and allowing cool brown dwarfs to be readily identifiable. 

Hundreds of potentially ultra-cool brown dwarf candidates have been identified using the WISE All-Sky Data. A detailed description of this search is described in \cite{kirkpatrick11} and \cite{kirkpatrick12}, where, along with \cite{mainzer11}, \cite{wright2011}, \cite{burgasser2011}, \cite{cushing11}, \cite{gelino2011}, and Mace et al. (in prep.), follow-up and preliminary analyses have been presented.  With the relatively shallow depth of WISE, a majority of these ultra-cool brown dwarf candidates either have very faint $W1$ detections or were simply not detected in $W1$, resulting in $W1-W2$ color limits. The {\it{Spitzer}} Infrared Array Camera (IRAC) 3.6 and 4.5 $\mu$m bands \citep{fazio04}, sometimes referred to as $ch1$ and $ch2$ are very similar to the $W1$ and $W2$ filters in WISE. This makes IRAC an excellent instrument for deeper follow-up observations (see Figure 2 in \citealt{mainzer11}).

In this paper, we present the photometric properties from IRAC follow-up of a large sample ($\sim$1500) of WISE-selected brown dwarf candidates (see Section 4.1). Of these, 182 have been spectroscopically classified, with classifications ranging from a single M dwarf to 12 examples of the latest and coldest spectral type, Y (\citealt{kirkpatrick08}; \citealt{cushing11}; \citealt{kirkpatrick12}). We also present new, ultra-cool brown dwarf candidates discovered using the  {\it{Spitzer}} data alone in an effort to identify fainter, widely separated companions to the WISE sources. As a control sample we also identify field brown dwarf candidates from a {\it{Spitzer}} campaign to follow-up WISE-selected hyper-luminous infrared galaxy (HyLIRG) candidates (see Section 4.2), as this enables us to assess whether the {\it{Spitzer}}-selected brown dwarf companion candidates are more likely to be truly associated with the WISE-selected brown dwarf candidates. We release a catalog for 906 WISE-selected HyLIRG candidates. In addition to these two extremely rare population of astrophysical sources, we release a photometric catalog of $\sim$ 7.33 $\times 10^5$ sources detected in the 1000's of {\it{Spitzer}} Astronomical Observation Requests (AOR's) in these follow-up campaigns. In $\S$2 we briefly describe the observations. We discuss the photometry and source detection in $\S$3. In $\S$4 we describe the catalogs presented in this release. We discuss the new brown dwarf candidates discovered by this analysis in $\S$5 and summarize the paper in $\S$6. All magnitudes are given in the Vega system unless otherwise noted. 

\section{Observations}

Warm {\it{Spitzer}} observations were carried out at 3.6 and 4.5 $\mu$m under {\it{Spitzer}} programs 70062, 70162, 80033, and 80109. Both the brown dwarf and HyLIRG candidate field  
observations used a 5-point dither pattern with 30 second exposures per pointing in each IRAC band. IRAC has a $5'$x$5'$ field of view with 1{\farcs}2 pixels. For this analysis we utilized the post-BCD {\it{Spitzer}} pipeline images, which have been resampled onto 0.6$''$ pixels. The AORs were executed between June 2010 and May 2012 and comprise a total of 1564 brown dwarf and 919 extragalactic follow-up observations. These observations have been described in detail in \cite{kirkpatrick11} and \cite{eisenhardt2012}. The images for this analysis were reduced with two different versions of the {\it{Spitzer}} reduction pipeline, 18.18 and 19.1, with the only change between the two versions being an update to the masking process which includes knowledge of latents going back to previous AOR's. None of these changes affect our reduction and processing techniques.

\section{Detection and Photometry}

Source detection and photometry were carried out using SExtractor \citep{bertin96}. We constructed a $ch2$ selected catalog by using the dual-image mode capabilities of SExtractor. In dual-image mode, sources are detected and their centroids and apertures are defined in one image and subsequently the photometry is measured in another image using those predetermined apertures and source centers. As in \cite{eisenhardt2010}, the choice of using $ch2$ to construct this catalog was motivated by the fact that cool brown dwarfs are generally brighter in $ch2$ than $ch1$. Photometry was measured in 6{\farcs}0 diameter apertures and corrected using the calibration values determined by the IRAC instrument team, -0.133 and -0.113 for $ch1$ and $ch2$, respectively. We use the IRAC zero points provided by the {\it Spitzer} Science Center (SSC), which are 18.789 and 18.316 Vega magnitudes for $ch1$ and $ch2$, respectively. SExtractor was configured to define a source as a set of twenty or more connected pixels, each lying 1.0$\sigma$ above the background. We utilize the post-BCD coverage maps to provide an exposure flag (COVCH\#) for all sources in the full catalog, with COVCH\#  equal to the number of frames going into the post-BCD stack at that position. We also provide the SExtractor parameter CLASS\_STAR to aid in separating point sources from extended sources. We compared the SExtractor aperture magnitudes for our brown dwarf candidates to an independent measurement of aperture photometry as presented in Kirkpatrick et al. (2011,2012).  This independent photometry tool was written in IDL using both public scripts from IDL Astronomy User's Library\footnote{http://idlastro.gsfc.nasa.gov/} and proprietary code created specifically for this task. In short, this IDL code centroids on the known position of the brown dwarf candidate, obtains the aperture photometry of the source, and applies the appropriate aperture corrections as provided by the SSC.  We find a mean offset $\Delta$$ch2$ $\sim$0.001 mag and a standard deviation sigma ($\Delta$$ch2$) $\sim$0.04 mag between the two codes, indicating that both are behaving reasonably.

\section{The Catalogs}

There are three separate catalogs presented in this paper. The first, described in \S4.1 and Table 1, is a list of {\it Spitzer} and WISE photometry for 1510 WISE-selected brown dwarf candidates targeted in {\it Spitzer} programs 70062, 80109. The second, described in \S4.2 and Table 2, is a list of {\it Spitzer} and WISE photometry for 906 WISE-selected HyLIRG candidates (hereafter, ``W1W2-dropouts," \citealt{eisenhardt2012}) that were the targets for {\it Spitzer} program 70162 and 80033. The third, described in \S4.3 and Table 3, is a list of {\it all} (COV $\ge$ 3) sources found on the {\it Spitzer} IRAC images for the three programs above.

\subsection{The Brown Dwarf Catalog}

Brown dwarf candidates were selected from the WISE source databases using the color criteria discussed in \cite{kirkpatrick11} and \cite{kirkpatrick12}. To summarize, our main selection used a color cut chosen to select cold brown dwarfs with types $\ge$T5. A color cut of $W1-W2 > 1.5$ mag was used with WISE internal source lists from early processing runs and $W1-W2 > 2.0$ mag was used for later processing, specifically using the combination of the WISE All-Sky Source Catalog and WISE All-Sky Reject Table that are contained in the WISE All-Sky Data Release (\citealt{cutri2012}). We also performed another cut with a much more relaxed color criterion ($W1-W2 > 0.4$ mag) to select bright (W2 S/N $>$ 30), nearby brown dwarfs with types $\ge$L5. Other constraints, including the lack of a positional match in earlier-epoch all-sky surveys (2MASS, DSS1, DSS2), cuts on the $W2-W3$ color, stipulations on the reduced $\chi^2$ value from Point Spread Function photometric fitting, etc.\ were also imposed, as discussed in Kirkpatrick et al. (2011) and Kirkpatrick et al. (2012).

Most of the selected sources were scheduled for {\it Spitzer} $ch1$ and $ch2$ observations as part of programs 70062 and 80109. As of late-May 2012, a total of 1564 unique candidates had been observed in both IRAC bands. We were not able to identify a {\it Spitzer} counterpart for 54 of these candidates, which were therefore deemed to be spurious sources in WISE. Further inspection revealed that these spurious sources fell into two categories: (a) cosmic rays that bled through to the final images in the WISE coaddition process due to multiple cosmic ray hits and low frame coverage, and (b) WISE sources within nebulous patches (or nebulae themselves) rather than nearby brown dwarfs. For the remaining 1510 candidates, however, {\it Spitzer} counterparts were easily detected in both $ch1$ and $ch2$ bands. These 1510 sources comprise the WISE-selected Brown Dwarf Photometric Catalog, as described in Table 1.

The WISE and {\it Spitzer} photometry for each source in this catalog is graphically illustrated in Figure 1. This figure plots the {\it Spitzer}/IRAC $ch1-ch2$ color as a function of WISE $W1-W2$ color. The 1510 sources from the catalog are shown by black dots. A total of 166 of these have spectroscopic confirmation from \cite{mainzer11}, \cite{burgasser2011}, \cite{kirkpatrick11}, \cite{cushing11}, \cite{kirkpatrick12} and Mace et al. (in prep), and are plotted with colored symbols, as indicated in the legend of the plot. The coldest brown dwarfs, the Y dwarfs, populate the reddest locus. The location of our latest type Y dwarf, WISE J182831.08+265037.7, which was targeted in a different {\it Spitzer} program (program 551; Mainzer, PI), is shown by the magenta octagon. As further illustrated in Figure 8 of \cite{kirkpatrick12}, the {\it Spitzer} $ch1-ch2$ color for WISE 1828+2650, typed as $\ge$Y2, is not as red as objects typed slightly earlier, at Y1. 

Many excellent brown dwarf candidates in the catalog and on this diagram still lack spectroscopic data. Of the five objects in Figure 1 with colors of $ch1-ch2 > 2.8$ mag, three are promising Y dwarf candidates -- WISE J064723.23$-$623235.5 ($ch1-ch2$ = 2.87, $W1-W2 > 3.77$), WISE J082507.35+280548.5 ($ch1-ch2$ = 2.96, $W1-W2 > 3.75$), and WISE J220905.73+271144.00 ($ch1-ch2$ = 2.98, $W1-W2 > 3.68$). All three are extremely faint in the $J$ and $H$ passbands, so we are awaiting spectroscopic confirmation using the WFC3 grisms onboard the {\it Hubble Space Telescope}. The other two sources, WISE 035358.23+375458.5 ($ch1-ch2$ = 3.00, $W1-W2 = 3.52$) and WISE J141127.45$-$612925.6 ($ch1-ch2$=3.42, $W1-W2 > 3.35$), are located within nebular patches on the {\it Spitzer} images and appear to be reddened or embedded objects rather than brown dwarfs.

These last two sources highlight an important point: spectroscopic confirmation of sources is vital to establishing candidates as bona fide brown dwarfs. Sixteen of our targets, shown by the yellow dots in Figure 1, have been shown spectroscopically {\em not} to be brown dwarfs (Mace et al. in prep).
These scatter widely over the diagram, proving that there is contamination at all colors. One interesting feature can be seen in the lower left quadrant of the diagram where our L and early-T dwarf candidates lie. These objects have bright WISE and {\it Spitzer} magnitudes, so the colors of these objects are well measured. These high-quality measurements enable us to split objects with $W1-W2 < 1.5$ mag and $ch1-ch2 < 1.0$ mag into two tracks. The lower track is replete with spectroscopically confirmed L and early-T dwarfs and primarily have 2MASS detections; the upper track has fewer confirming observations and generally no 2MASS detections, but objects with spectra here are typically not found to be brown dwarfs. Users of this catalog are cautioned that the sources here are candidates only, our selections were fairly liberal, so contamination by other types of sources is expected.

\subsection{The W1W2-Dropout Catalog}

\cite{eisenhardt2012} have identified 
a rare population of objects termed ``$W1W2$-dropout" galaxies, 
because they are faint or undetected in $W1$ and $W2$ but well detected in $W3$ or $W4$. 
Optical spectroscopy shows that most of these objects have redshift $z > 1.6$  \citep{eisenhardt2012}, 
which in combination with sub-millimeter follow-up detections results in bolometric luminosities 
$> 10^{13} L_\odot$ and in some cases $> 10^{14} L_\odot$ \citep{wu2012}. 
This qualifies them as hyper-luminous infrared galaxies or HyLIRGs, the other
primary science objective for WISE (the first being cool brown dwarfs, as noted in the introduction).
The sub-millimeter data show their luminosities are dominated by dust with temperatures more than
twice as high as other infrared luminous populations, which leads \cite{wu2012} to refer to them
as hot dust obscured galaxies or ``hot DOGs". A closely related population shows a high incidence (1/3) of very extended ($>$ 30 kpc) Ly-$\alpha$ emission
or Ly-$\alpha$ blobs, which normally are found in less than 1\% of galaxies, as described in \cite{bridge2012}.
The extreme luminosities, hot dust, and prevalence of Ly-$\alpha$ blobs in this population 
suggests they may be a rare phase in the co-evolution of galaxies and their central supermassive black holes. 

Approximately 1,000 $W1W2$-dropouts have been identified in the WISE catalog, 
and because of their faintness in W1 and W2, 919 have been followed up in the
{\it Spitzer} IRAC $ch1$ and $ch2$ bands.
Of the 919 sources, 910 were selected in {\it Spitzer} cycle 7,
9 were added in cycle 8, and 13 failed to meet the
SExtractor detection thresholds described in section 3.

The  IRAC $ch1$ and $ch2$ bands are similar in wavelength to W1 and W2 
but can easily reach much greater depths.  These bands sample near-infrared 
wavelengths in the rest-frame at the typical $z \sim 2$ of $W1W2$-dropouts, providing
a measure of the stellar population in the galaxies.  Figure 1 of \cite{eisenhardt2012}
shows the resulting distribution of $ch1 - ch2$ colors converted to $W1 - W2$. Here we present (in Table 2) 
{\it Spitzer} and WISE photometry for the 906 sources detected in the {\it Spitzer} data by SExtractor. 

The bulk of the $ch1 - ch2$ colors lie well above $W1 - W2 = 0.8$, which indicates the source has a high probability of being an Active Galactic Nucleus (AGN), as \cite{stern2012} 
show. The optical spectra also often show AGN signatures  \citep{wu2012} , and it appears likely that
the bulk of their extraordinary luminosity is powered by accretion onto supermassive
black holes.  As shown in  \cite{eisenhardt2012}, the unusually high ratio of $W3$ and $W4$ 
emission relative to $ch1$ suggests a higher ratio of supermassive black hole mass to
stellar mass than is found in local galaxies, implying the black holes grow before
their stellar populations.  This sequence is unexpected in the most frequently
discussed scenarios of galaxy and AGN feedback (e.g. \citealt{hopkins2012}).

\subsection{The Complete Photometric Catalog}

We provide a catalog of the $\sim$ 7.33 $\times 10^5$ sources found in the {\it{Spitzer}} fields (COVCH1 $\ge$ 3 and COVCH2 $\ge$ 3) whose properties are 
described in Table 3. We provide WISE and 2MASS  photometry from the WISE All-Sky source catalog for sources that fall within 2$''$ of a {\it Spitzer} source. In addition, we provide optical photometry for {\it Spitzer} sources that fall within 2$''$ of a source in the USNO-B catalog. Table 4 summarizes the statistics that resulted from matching between these catalogs for two different samples. Sample A comprises sources with the deepest imaging and most reliable photometric measurements available (COVCH1 = 5 and COVCH2 = 5). Sample B represents sources with slightly shallower imaging (COVCH1 $\ge$ 3 and COVCH2 $\ge$ 3) and thus slightly lower photometric reliability, though still reasonably good enough for most scientific purposes. 

To characterize the photometric sample, we provide color-color and color-magnitude diagrams in Figure 2.  The left plot of Figure 2 is similar to Figure 1, but includes all sources from Sample A having WISE measurements (a total of $\sim 10^5$ sources) rather than just the WISE brown dwarf candidates. We also show these colors for the 166 spectroscopically confirmed brown dwarfs, plotted with colored symbols as indicated in the legend of the plot. Few field sources in the {\it Spitzer} images have colors as extreme as the latest T and Y dwarfs.

In the center plot of Figure 2 we show $ch1 - ch2$ versus $ch2$ mag. The limiting magnitude is $ch2 \sim 18.0$ mag, which is $\sim$ 2.5 
mag fainter then the WISE-selected candidates and comparable to the depth of the IRAC Shallow Survey (Eisenhardt et al 2004), which had similar exposure time. 
We see that the majority of the WISE-selected brown dwarf candidates occupy a diffuse region in this color-magnitude domain, mainly with $ch2 < 15.5$ and $ch1 - ch2 > 0.7$.  

It is challenging to separate Galactic sources from extragalactic sources using the $ch1$ and $ch2$ filters in the IRAC camera alone. In the rightmost plot of Figure 2 we show $ch1 - ch2$ vs $ch2 - W3$. Here we can see the benefit of including photometric measurements at $12\mu$m. The additional color information helps separate the Galactic and extragalactic red ($ch1 - ch2 > 1$) populations with cool brown dwarfs having $ch2 - W3 < 3$ while the extragalactic population has $ch2 - W3 > 3$. A similar separation is shown in Figure 1 of \cite{eisenhardt2010}.

\section{New Brown Dwarf Candidates Selected From {\it{Spitzer}}}

We present new brown dwarf candidates discovered from a blind search using the following search criteria $ch1 - ch2 \ge 1.5$, 
$ch2 \le 17.0$ mag and coverage equal to 5 in both IRAC channels (sample A from Table 4). 
As noted by \cite{eisenhardt2010}, this choice of color is expected for sources with spectral types later than T7 and the magnitude cut allows for only robustly measured sources to be inspected. This search resulted in 666 candidates from the brown dwarf fields and 247 candidates from the extragalactic fields. We then used the Basic Calibrated Data (BCDs) to inspect the individual frames to remove any spurious sources or sources which appeared to be contaminated by cosmic rays. After also removing sources which were the WISE-selected targets, we were left with 13 new ultra-cool brown dwarf candidates: seven in the WISE brown dwarf fields, and six in the WISE HyLIRG fields. Table 5 lists these 13 candidates, and Multi-wavelength finder charts ($1' \times 1'$) for them are presented in Figures 3, 4, and 5. The finder charts show that all of these candidates lack detections at shorter wavelengths and 12 $\mu$m, as expected for brown dwarf candidates. However, the {\it Spitzer} images are relatively deep compared to the observations at shorter (DSS,2MASS) and longer wavelengths (WISE). The difference in depth between these sets of observations makes it difficult to exclude certain types of extra-galactic objects such as AGN. 

For comparison \cite{eisenhardt2010} found 32 similar brown dwarf candidates in the deeper Spitzer Deep Wide Field Survey (SDWFS) (\citealt{ashby09}) covering 10 square degrees, but selecting candidates to a fainter limit, $ch2 \le 18.5$. Of these 32 SDWFS brown dwarf candidates, 7 had $ch2 \le 17.0$. After accounting for the coverage depth of 5 criterion, the 2483 IRAC pointings searched here cover a comparable area to SDWFS, suggesting a somewhat higher surface density of brown dwarfs candidates than in SDWFS. All of the SDWFS candidates with $ch2 \le 17.0$ were among those rejected as probable AGN because they had $ch2 - ch4 > 2$. An analogous AGN rejection criterion using the $ch2 - W3$ color is not possible here because the $W3$ depth is not adequate, so it is plausible that many of the 13 brown dwarf candidates identified here are AGN. If so, it is possible that the somewhat higher surface density is attributable to AGN clustering. 


Determining whether any of the new candidates could potentially be widely separated companions can be achieved in two ways. First, we check the surface density of our brown dwarf candidates in the control sample versus that in the brown dwarf fields. We find a surface density of (1.5 $\pm 0.6$ deg$^{-2}$) in the control sample and (1.0 $\pm 0.4$ deg$^{-2}$) in the brown dwarf fields. The errors are calculated based on simple Poisson statistics. The fact that the control field has a larger surface density suggest that the candidates found are unlikely to be physically associated companions. Second, we can determine the spectrophotometric distances to both the primary WISE-selected source and the new {\it Spitzer}-selected candidate. We estimate the spectral type by using Figure 11 from \cite{kirkpatrick11} and the derived $ch1-ch2$ color. We then estimate the photometric distances listed in Table 5 using the apparent magnitude and the absolute magnitude, where the latter is estimated using the absolute magnitude versus spectral type relationship in Figure 13 of \cite{kirkpatrick12}.

Most of the new candidates do not appear to be physically associated with the WISE-selected targets, but are more distant background sources. Assuming the WISE-selected candidates are indeed brown dwarfs, they reside at $\sim$ 20 pc, while the majority of the new {\it Spitzer}-selected candidates are at $\simgt $ 40 pc. There is only one case for which both the primary and the candidate may be physically associated, J052439.3+464631.2, where both sources have estimated distances of $\sim$ 20 pc. The estimated spectral types for these two sources are T6 and T9. The angular separation of 83.26$''$ for this system corresponds to a physical separation of at least 1600 AU at 20 pc. 

The probability of discovering widely separated companions rests on the probability that such systems exist, which is itself a complicated function of mass, age, environment, and formation mechanisms. An in depth review on multiplicity in very low mass systems is given in \cite{burgasser07} and references therein. 
These brown dwarf primaries have low mass and thus small gravitational potential wells. This is well summarized in Figure 6 of \cite{burgasser07}, where they plot the separation of binaries as a function of total mass for a wide variety of known systems. 
Brown dwarf binaries tend to have small separations with higher binding energies.
Systems with lower binding energies tend to be disrupted, 
primarily due to outside perturbers or gravitational instabilities, 
which translates to almost no old brown dwarf binaries having been found 
with separations greater than 20 AU. There are only a handful of low-mass binaries with separations greater than 20 AU, three are relatively young and thus have not had enough time for collisional disruption. Another system is DENIS 0551-4443AB \citep{billeres05}, which is thought to be relatively old and is the kind of widely separated  system for which we were searching. \cite{billeres05} suggest that such a system is fragile, and it would not have survived a close encounter with a third body. Furthermore, its existence demonstrates that some very low mass stars/brown dwarfs form without ejection from a multiple system, or any other strong dynamical interaction. Other widely separated multiple systems have been discovered since \cite{burgasser07}, including the NLTT20346 + 2MASS J0850359+105716 system (\citealt{faherty11}). Is it possible that J052439.3+464631.2 is also an old, widely separated binary like these?

The nature of J052439.3+464631.2 can be probed further by considering the environment, magnitudes, and colors
of the individual objects in this system. Using the WISE All-Sky Image Archive we investigate a region $\sim$ 5$'$ on a side centered on the new candidate and make the following observations. First, this system resides near the Galactic Plane which makes it more difficult to interpret given the high source density and complex nebulosity. The pair resides at the edge of a 12$\mu$m bright nebula, thus suggesting that rather than a widely separated T6-T9 binary, these are more likely reddened or embedded objects and not brown dwarfs. Another possibility is that they could be physically associated T6 and T9 brown dwarfs, though not in a binary system, 
i.e. forming in the same stellar cocoon and around the same time. Spectroscopic observations of these candidates should be made to disentangle these scenarios. 

Other than J052439.3+464631.2, this search yielded no plausible widely separated companions, therefore suggesting that very low mass ($>$T7) brown dwarf binaries
with separations of 20 to 2000 AU are extremely rare. 

\section{Summary}

This paper 
summarizes initial {\it Spitzer} follow-up and analysis of 
some of the rarest astrophysical objects discovered to date. We present, for the first time, a very large sample of WISE-selected ultra-cool brown dwarf candidates. Of these, 184 have spectroscopic data with $\sim$ 92\% of them confirmed as bona fide ultra-cool brown dwarfs and of which $\sim$ 83\% have spectral types later than T5. We also present 13 new, though more distant, ultra-cool ($>$T7) brown dwarf candidates discovered using the {\it Spitzer} data.  
Only one possible widely separated companion system was found,
suggesting that widely separated, cold brown dwarf binaries 
are extremely rare. To date, the WISE team has only followed up a small percentage ($\sim$ 12\%) of the total WISE-selected sample, providing follow-up opportunities by the astrophysical community on a host of interesting discoveries yet to be made. Given the wide range of (sub)stellar spectral types it probes, the sample will allow for a multitude of future statistical studies. In addition, this sample will provide excellent follow-up sources for the {\it James Webb Space Telescope} (JWST). 

There are several avenues of investigation yet to be fully explored which we summarize as follows. Detailed proper motion measurements can be made with the current data available (WISE + {\it Spitzer}), though long term monitoring will allow for full kinematics and robust distances to be determined. Several large parallax programs are currently underway monitoring a large number of the spectroscopically verified sources, though many remain to be monitored. Long term monitoring will also shed critical insight and constraints on variability in brown dwarfs, two examples being the study of atmospheric meteorology and possible detection of transits by planetary companions. Using the difference between the WISE and {\it Spitzer} filter profiles will allow one to explore differences in the underlying spectra at these wavelengths. With no all-sky deep surveys planned at these wavelengths in the foreseeable future, this is the quintessential brown dwarf sample because it comprises the closest and brightest brown dwarfs possible.

In addition to presenting these two very rare populations of astrophysical sources, we release a catalog of $\sim$ 7.33 $\times 10^5$ objects detected in the {\it Spitzer} observations, including multi-wavelength (2MASS, USNO-B, and WISE) photometry for a subsample. Providing multi-wavelength photometry for hundreds of thousands of astrophysical sources will allow for a number of scientific studies to be conducted, from fully characterizing the mid-IR properties of the coldest brown dwarfs to a more complete characterization of the most IR luminous galaxies in the universe.

This publication makes use of data products from the {\it Wide-Field Infrared Survey Explorer}, which is a joint project of the University of California, Los Angeles, and the Jet Propulsion Laboratory/California Institute of Technology, funded by the National Aeronautics and Space Administration (NASA). This publication also makes use of observations made with the {\it Spitzer Space Telescope}, which is operated by the Jet Propulsion Laboratory, California Institute of Technology, under a contract with NASA. 
This work is also based in part on observations made with the NASA/ESA {\it Hubble Space Telescope}, obtained at the Space Telescope Science Institute, which is operated by the Association of Universities for Research in Astronomy, Inc., under NASA contract NAS 5-26555. Some of the spectroscopic classifications presented herein were obtained at the W. M. Keck Observatory, which is operated as a scientific partnership among the California Institute of Technology, the University of California and the National Aeronautics and Space Administration. This publication also makes use of data products from 2MASS. 2MASS is a joint project of the University of Massachusetts and the Infrared Processing and Analysis Center/California Institute of Technology, funded by the National Aeronautics and Space Administration and the National Science Foundation. This research has made use of the NASA/IPAC Infrared Science Archive (IRSA), which is operated by the Jet Propulsion Laboratory, California Institute of Technology, under contract with NASA.

\newpage

\begin{table*}[h!tb!]
\caption{Brown Dwarf Photometric Catalog Parameters}
\begin{center}
\begin{tabular}{lll}\hline\hline
Parameter & Example value & Description \\\hline
   DESIGNATION   & 205628.91+145953.2 & WISE sexigesimal designation \\
   SPTYPE          &  Y0  & Near-IR Spectral Type \\
   RA              &        314.12046 & Right ascension in decimal degrees J2000\\
   DEC            &    14.998111 & Declination in decimal degrees J2000\\
   CH1MAG          &   15.873& 3.6 $\mu$m Vega magnitude\\
   CH1ERR         &   0.017& 3.6 $\mu$m Vega magnitude error\\
   CH2MAG          &    13.914& 4.5 $\mu$m Vega magnitude\\
   CH2ERR          &  0.003& 4.5 $\mu$m Vega magnitude error\\
   W1MPRO           &   18.253& WISE 3.4 $\mu$m profile-fit magnitude \\
   W1SIGMPRO           &     0.000& WISE 3.4 $\mu$m profile-fit  magnitude error \\
   W1SNR          &  -0.300& WISE 3.4 $\mu$m signal-to-noise ratio \\
   W2MPRO           &     13.928& WISE 4.6 $\mu$m profile-fit  magnitude \\
   W2SIGMPRO           &     0.046& WISE 4.6 $\mu$m profile-fit  magnitude error \\
   W2SNR          &       23.400& WISE 4.6 $\mu$m signal-to-noise ratio \\
   W3MPRO          &       12.003&WISE 12.0 $\mu$m profile-fit  magnitude \\
   W3SIGMPRO           &     0.270&WISE 12.0 $\mu$m profile-fit  magnitude error \\
   W3SNR         &      4.000&WISE 12.0 $\mu$m signal-to-noise ratio \\
   W4MPRO          &        8.781&WISE 22.0 $\mu$m profile-fit  magnitude \\
   W4SIGMPRO           &          NaN& WISE 22.0 $\mu$m profile-fit  magnitude error \\
   W4SNR           &         0.800&WISE 22.0 $\mu$m signal-to-noise ratio \\
\end{tabular}
\tablecomments{When W\#SIGMPRO is either 0.000 or NaN it is the same as null in the WISE All-sky Source Catalog, see http://wise2.ipac.caltech.edu/docs/release/allsky}
\end{center}
\end{table*}

\begin{table*}[h!tb!]
\caption{W1W2-dropout Photometric Catalog Parameters}
\begin{center}
\begin{tabular}{lll}\hline\hline
Parameter & Example value & Description \\\hline
   DESIGNATION   & 000025.1+420708.5 & WISE sexigesimal designation \\
   RA              &        0.10475330 & Right ascension in decimal degrees J2000\\
   DEC            &    42.119005 & Declination in decimal degrees J2000\\
   CH1MAG          &   16.759& 3.6 $\mu$m Vega magnitude\\
   CH1ERR         &   0.039& 3.6 $\mu$m Vega magnitude error\\
   CH2MAG          &    15.837& 4.5 $\mu$m Vega magnitude\\
   CH2ERR          &  0.022& 4.5 $\mu$m Vega magnitude error\\
   W1MPRO           &   17.912& WISE 3.4 $\mu$m profile-fit magnitude \\
   W1SIGMPRO           &     0.241& WISE 3.4 $\mu$m profile-fit  magnitude error \\
   W1SNR          &  4.500& WISE 3.4 $\mu$m signal-to-noise ratio \\
   W2MPRO           &     16.198& WISE 4.6 $\mu$m profile-fit  magnitude \\
   W2SIGMPRO           &     0.176& WISE 4.6 $\mu$m profile-fit  magnitude error \\
   W2SNR          &       6.200& WISE 4.6 $\mu$m signal-to-noise ratio \\
   W3MPRO          &       11.474&WISE 12.0 $\mu$m profile-fit  magnitude \\
   W3SIGMPRO           &     0.114&WISE 12.0 $\mu$m profile-fit  magnitude error \\
   W3SNR         &      9.500&WISE 12.0 $\mu$m signal-to-noise ratio \\
   W4MPRO          &        7.50&WISE 22.0 $\mu$m profile-fit  magnitude \\
   W4SIGMPRO           &          0.092& WISE 22.0 $\mu$m profile-fit  magnitude error \\
   W4SNR           &         11.800&WISE 22.0 $\mu$m signal-to-noise ratio \\
\end{tabular}
\tablecomments{When W\#SIGMPRO is either 0.000 or NaN it is the same as null in the WISE All-sky Source Catalog, see http://wise2.ipac.caltech.edu/docs/release/allsky}
\end{center}
\end{table*}

\begin{table*}[h!tb!]
\caption{Spitzer Photometric Catalog Parameters}
\begin{center}
\begin{tabular}{lll}\hline\hline
Parameter & Example value & Description \\\hline
   AOR\_KEY     &  r40819456 & Astronomical Observation Request \\
   SURVEY        &  BD   & Origin of {\it{Spitzer}} follow-up (BD or EXGAL) \\ 
   RA              &        332.35145 & Right ascension in decimal degrees (J2000)\\
   DEC            &    -27.565469 & Declination in decimal degrees (J2000)\\
   CH1MAG          &   14.183& 3.6 $\mu$m Vega magnitude\\
   CH1ERR         &   0.003& 3.6 $\mu$m Vega magnitude error\\
   CH2MAG          &    14.147& 4.5 $\mu$m Vega magnitude\\
   CH2ERR          &  0.003& 4.5 $\mu$m Vega magnitude error\\
   CLASS\_STAR\_CH1  &      0.930 & SExtractor star-galaxy separator measured 3.6 $\mu$m \\
   CLASS\_STAR\_CH2  &        0.950&  SExtractor star-galaxy separator measured at 4.5 $\mu$m \\
   COVCH1          &              5 & Total number of {\it Spitzer} frames at source position 3.6 $\mu$m \\
   COVCH2          &               5& Total number of {\it Spitzer}frames at source position 4.5 $\mu$m \\
   W1MPRO           &   14.192& WISE 3.4 $\mu$m profile-fit  magnitude \\
   W1SIGMPRO           &     0.032& WISE 3.4 $\mu$m profile-fit  magnitude error \\
   W1SNR          &  34.000& WISE 3.4 $\mu$m  signal-to-noise ratio \\
   W2MPRO           &     14.203& WISE 4.6 $\mu$m profile-fit  magnitude \\
   W2SIGMPRO           &     0.056& WISE 4.6 $\mu$m profile-fit  magnitude error \\
   W2SNR          &       19.400& WISE 4.6 $\mu$m  signal-to-noise ratio \\
   W3MPRO          &       12.661&WISE 12.0 $\mu$m profile-fit  magnitude \\
   W3SIGMPRO           &     NaN&WISE 12.0 $\mu$m profile-fit  magnitude error \\
   W3SNR         &      -0.3000&WISE 12.0 $\mu$m  signal-to-noise ratio \\
   W4MPRO          &       9.0360&WISE 22.0 $\mu$m profile-fit  magnitude \\
   W4SIGMPRO           &          NaN& WISE 22.0 $\mu$m profile-fit  magnitude error \\
   W4SNR           &         0.100&WISE 22.0 $\mu$m  signal-to-noise ratio \\
   JMAG            &      15.304& 2MASS $J$ magnitude \\
   JERR            &       0.055& 2MASS $J$ magnitude error \\
   HMAG            &         14.635& 2MASS $H$ magnitude \\
   HERR            &        0.063& 2MASS $H$ magnitude error\\
   KMAG            &          14.471& 2MASS $K_s$ magnitude \\
   KERR            &        0.083& 2MASS $K_s$ magnitude error \\
   BMAG           &      19.690&  USNO-B $B$2 magnitude\\
   RMAG            &         18.370&USNO-B $R$2 magnitude \\
\end{tabular}
\tablecomments{When W\#SIGMPRO is either 0.000 or NaN it is the same as null in the WISE All-sky Source Catalog, see http://wise2.ipac.caltech.edu/docs/release/allsky}
\end{center}
\end{table*}

\begin{table*}[h!tb!]
\caption{Photometric Catalog Statistics}
\begin{center}
\begin{tabular}{cccccc}\hline\hline
Sample&Coverage & \# w/{\it Spitzer} & \# w/WISE & \# w/2MASS & \# w/USNO-B \\\hline
A&=5 & 3.41$\times 10^5$ & 1.00$\times 10^5$ & 3.90 $\times 10^4$ & 7.75 $\times 10^4$ \\
B&$\ge$3 & 7.33$\times 10^5$ & 2.02$\times 10^5$ & 8.22 $\times 10^4$ & 1.63 $\times 10^5$
\end{tabular}
\end{center}
\end{table*}

\begin{sidewaystable}[h!tb!]
\caption{New Brown Dwarf Candidates}
\begin{center}
\begin{tabular}{lcccccccccccl}\hline\hline
Origin & \multicolumn{2}{c}{Candidate}  & \multicolumn{2}{c}{Candidate} &  \multicolumn{2}{c}{Primary} &Type$^a$ & Type$^b$& $d_{est}^c$& $d_{est}^d$ &Separation$^e$\\
& RA(J2000) & DEC(J2000)   & $ch2$  & $ch1-ch2$ & $ch2$&$ch1-ch2$&& & pc &pc &\\\hline
BD&02:39:59.8 &+70:40:20.4 &16.42$\pm$0.04& 1.91$\pm$0.16& 14.14$\pm$0.01& 0.78$\pm$0.01& T8& T5& 40.6& 21.8& 36.27$''$\\
EXGAL&03:02:21.1&02:16:09.8& 16.80$\pm$0.05 & 1.54$\pm$0.18& &&&&& \\
EXGAL&04:12:09.3&69:45:24.1& 16.94$\pm$0.06 & 1.65$\pm$0.22& &&&&& \\
BD& 05:24:39.3 & +46:46:31.2& 15.57$\pm$0.02& 2.29$\pm$0.11& 14.52$\pm$0.01& 1.03$\pm$0.01& T9& T6& 20.5& 21.5&83.26$''$ \\ 
EXGAL&08:21:51.3&23:45:17.1& 16.63$\pm$0.04 & 2.45$\pm$0.32& &&&&& \\
BD & 16:29:50.0& +25:45:25.9& 16.45$\pm$0.03& 1.68$\pm$0.12& 15.44$\pm$0.01& 1.06$\pm$0.03& T7.5& T6& 44.5& 32.8&139.13$''$\\ 
BD&16:53:42.8& -16:00:10.0& 16.91$\pm$0.08& 1.69$\pm$0.33& 14.61$\pm$0.01& 1.00$\pm$0.02& T7.5& T5.5& 54.92& 24.1&44.98$''$\\
EXGAL&20:38:38.1&-17:49:49.1&15.32$\pm$0.01 & 2.61$\pm$0.12& &&&&&  \\
EXGAL&21:41:45.5&01:20:14.5 & 15.95$\pm$0.02 & 2.27$\pm$0.19 & &&&&& \\
EXGAL&03:14:52.0&-53:52:41.9& 16.81$\pm$0.04 & 2.54$\pm$0.40 & &&&&& \\
BD&03:04:26.6& +43:31:10.9& 16.77$\pm$0.05& 1.93$\pm$0.22& 14.90$\pm$0.01& 1.22$\pm$0.02& T8& T7& 47.6& 23.0&21.38$''$\\
BD&18:22:38.6& -67:50:03.9& 16.66$\pm$0.04& 2.06$\pm$0.22& 15.12$\pm$0.01& 1.09$\pm$0.02& T9& T6.5& 33.9& 26.7&96.27$''$\\
BD$^f$&13:59:42.7& -23:32:35.6& 16.56$\pm$0.04& 1.61$\pm$0.16&&&T7 &&& \\\hline
\end{tabular}
\end{center}
$a$ Estimated brown dwarf type for new candidate \\
$b$ Estimated brown dwarf type for primary\\
$c$ Estimated distance to the new candidate, in parsecs\\
$d$ Estimated distance to the primary, in parsecs \\
$e$ Distance between new candidate and the primary, in arc seconds\\
$f$ primary $Spitzer$ target for this observation was not detected in the imaging, thus no information is given for the primary target \\
\end{sidewaystable}

\begin{figure*}[!t]
\begin{center}
\begin{tabular}{l}
\includegraphics[scale=1]{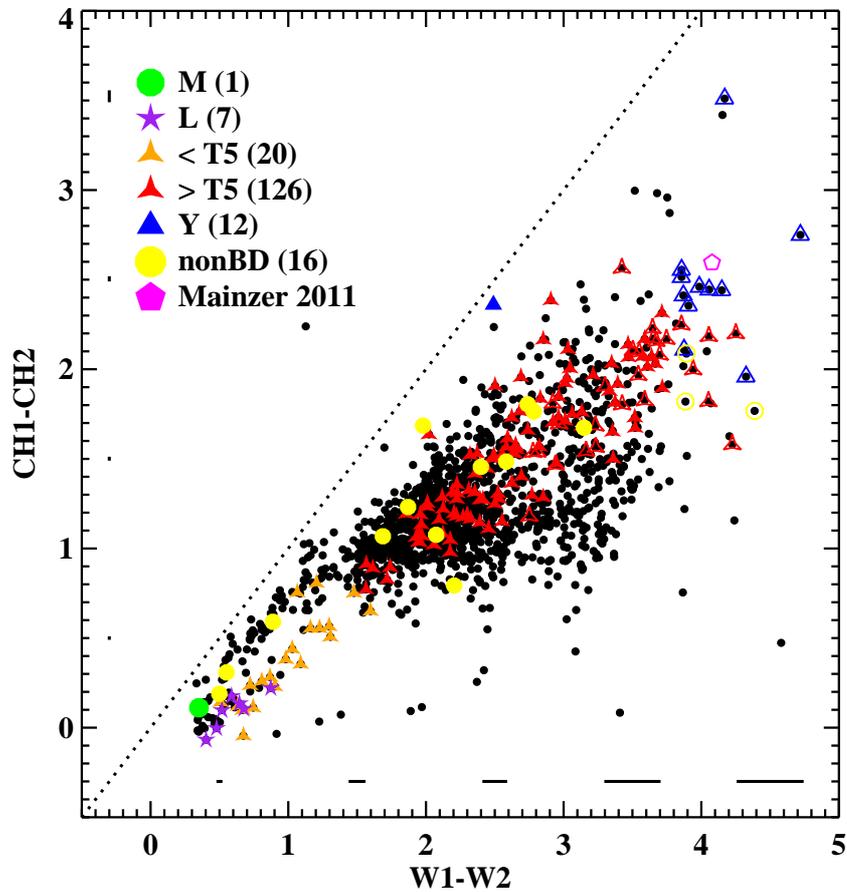}
\end{tabular}
\end{center}
\caption[CAPTION]{\label{fig:1} $Spitzer$/IRAC $ch1 - ch2$ as a function of $W1-W2$ color. Black points are observations of 1510 WISE-selected brown dwarfs detected by {\it Spitzer} . Those with spectroscopic classifications are overplotted in color, described in the legend. Open symbols denote color limits in $W1-W2$. We present the median uncertainties in each full magnitude bin of color for both axes and these are shown as the tick marks along the lower and left axes. The dotted line represents the 1-to-1 relationship between the axes. }
\end{figure*}

\begin{figure*}[!t]
\begin{center}
\begin{tabular}{l}
\includegraphics[scale=0.43]{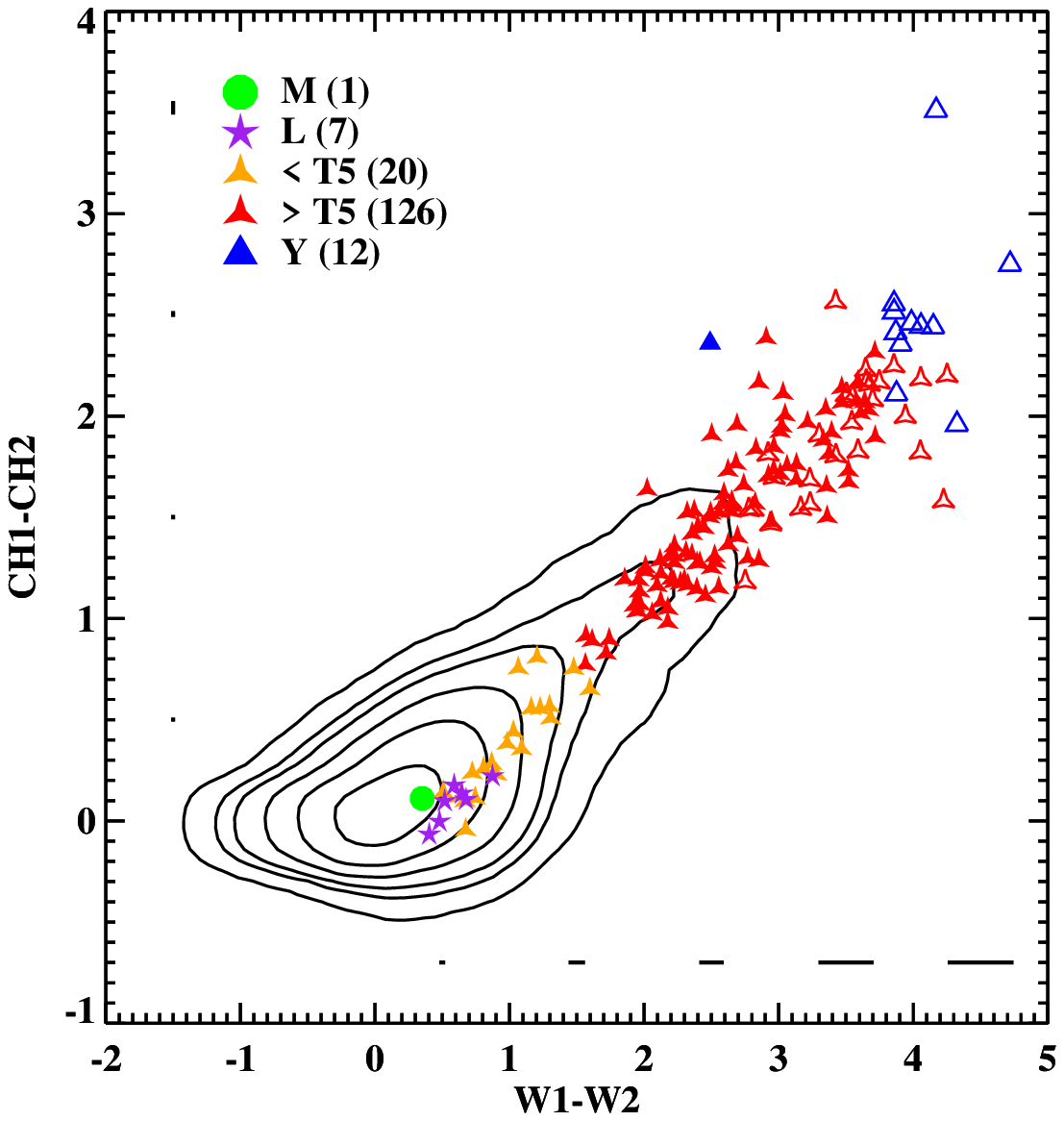}
\includegraphics[scale=0.43]{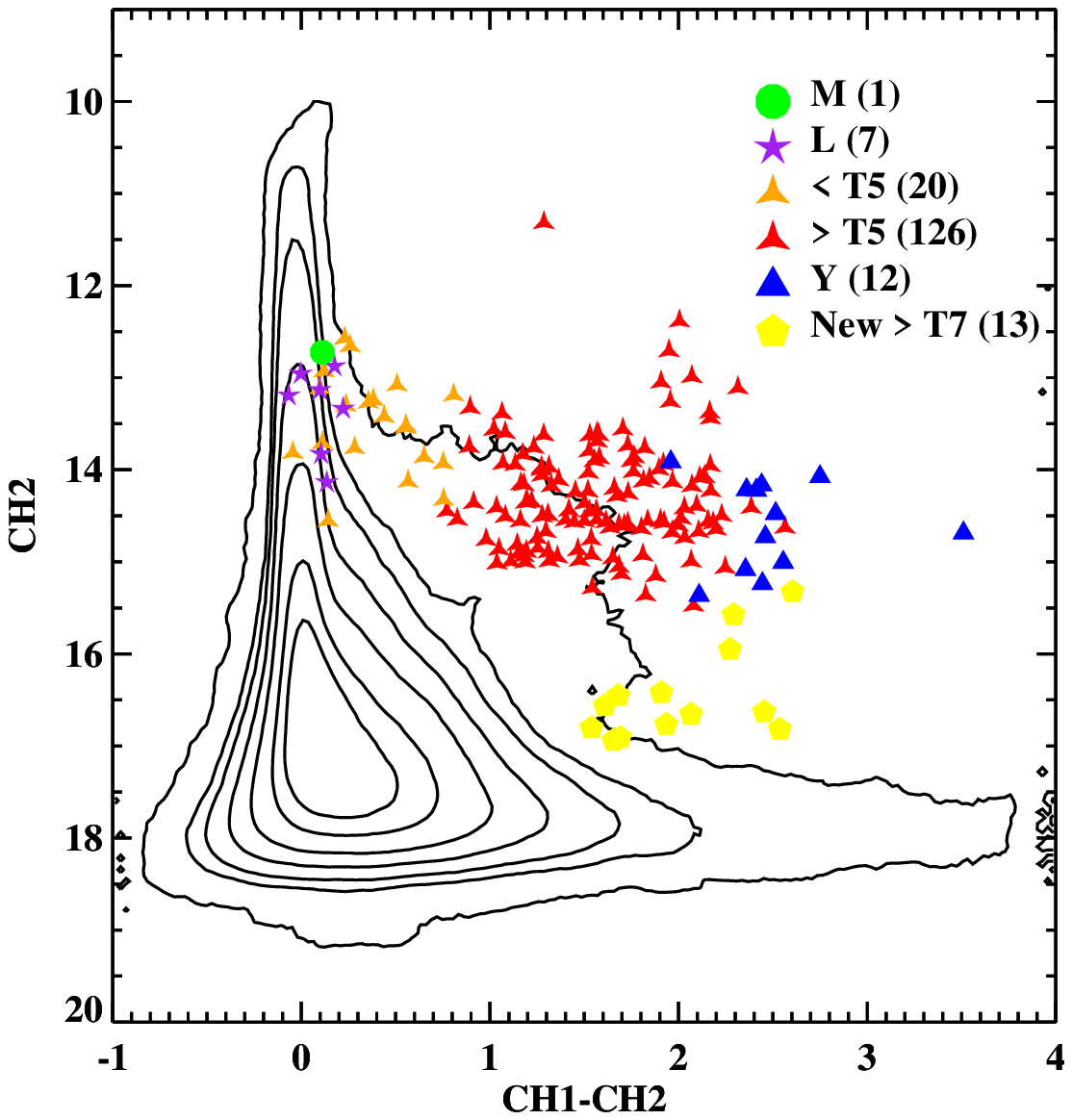}
\includegraphics[scale=0.43]{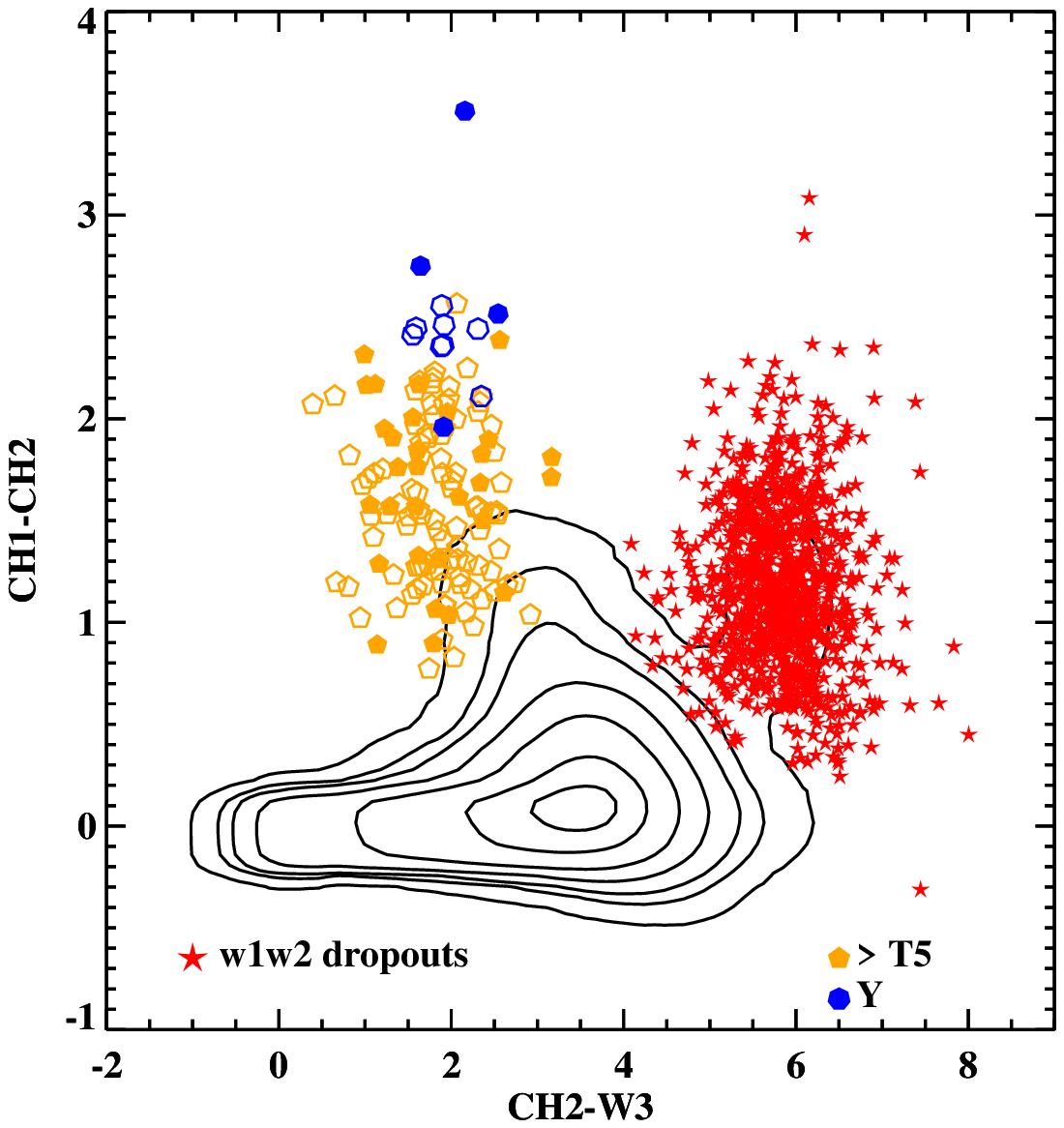}
\end{tabular}
\end{center}
\caption[CAPTION]{\label{fig:2} Left: $Spitzer$/IRAC $ch1 - ch2$ as a function of WISE $W1-W2$ color. Contours represent all sources with photometric measurements from these two instruments ($\sim 10^5$ sources). Objects with spectroscopic classification are shown by the colored points. Open symbols denote color limits in $W1-W2$. Center: $Spitzer$/IRAC $ch1 - ch2$ as a function of $ch2$ magnitude. Contours show $\sim 3.4\times 10^5$ sources (sample A in Table 4). Objects with spectroscopic classification are shown by the colored points. We also show the 13 newly discovered candidates as yellow circles. Right: $Spitzer$/IRAC $ch1 - ch2$ as a function of $ch2-W3$ color. Contours are similar to the left plot.  Brown dwarfs with spectroscopic classification greater than T5 are shown in orange (T) and blue (Y). We also present the WISE-selected HyLIRG candidates in red. Open symbols denote color upper limits.  The contours represent the 40, 60, 80, 90, 95, 97, and 99 percentiles of the distribution. Tick marks along the axes of the left panel are computed as discussed in Figure 1.}
\end{figure*}

\begin{figure*}[!t]
\begin{center}
\begin{tabular}{l}
\includegraphics[scale=0.75,angle=90]{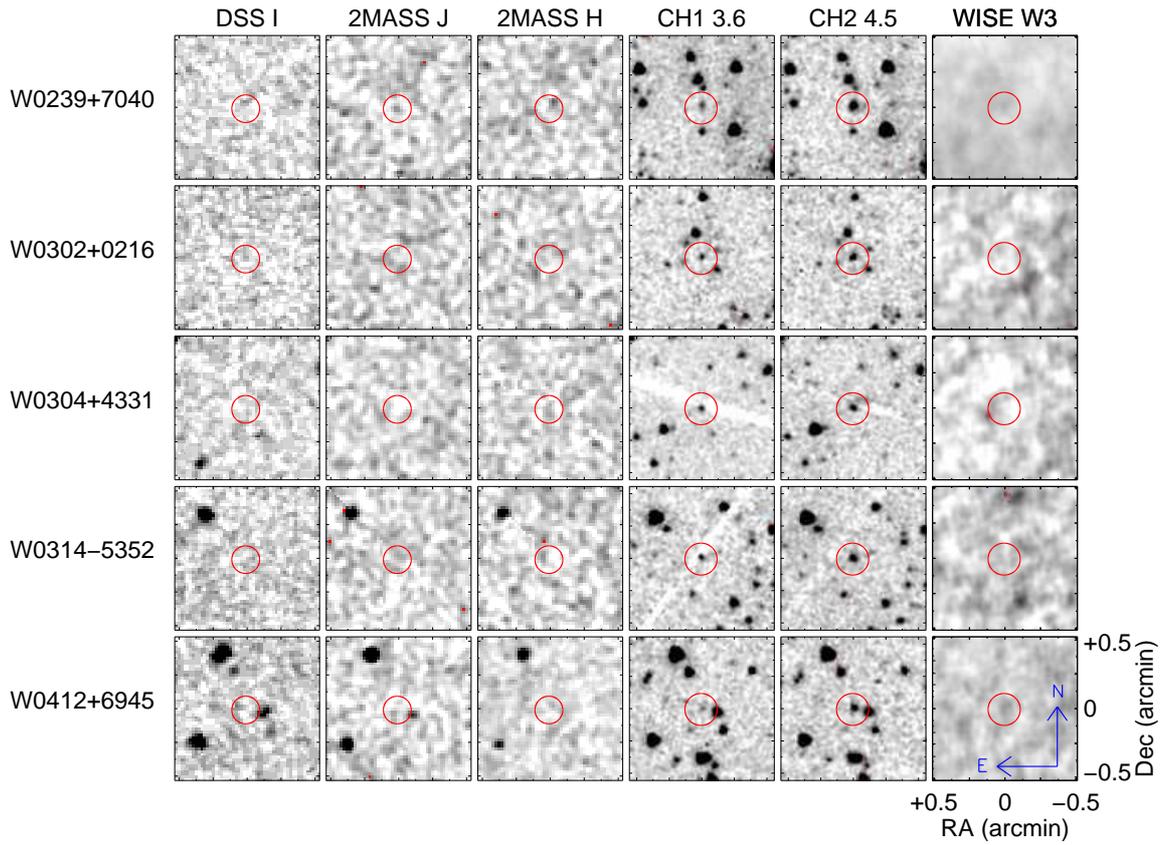}\\
\end{tabular}
\end{center}
\caption[CAPTION]{\label{fig:3} Multi-wavelength finder charts for the new ultra-cool brown dwarf candidates. We present $1\times1$ arcminute cutouts from various all-sky surveys alongside the {\it{Spitzer}} discovery images. From left to right, these cutouts are from the Digitized Sky Survey $I$ band, 2MASS $J$ and $H$, {\it{Spitzer}} $ch1$ and $ch2$ and WISE $W3$. The images are oriented with north up and east to the left.}
\end{figure*}

\begin{figure*}[!t]
\begin{center}
\begin{tabular}{l}
\includegraphics[scale=0.75,angle=90]{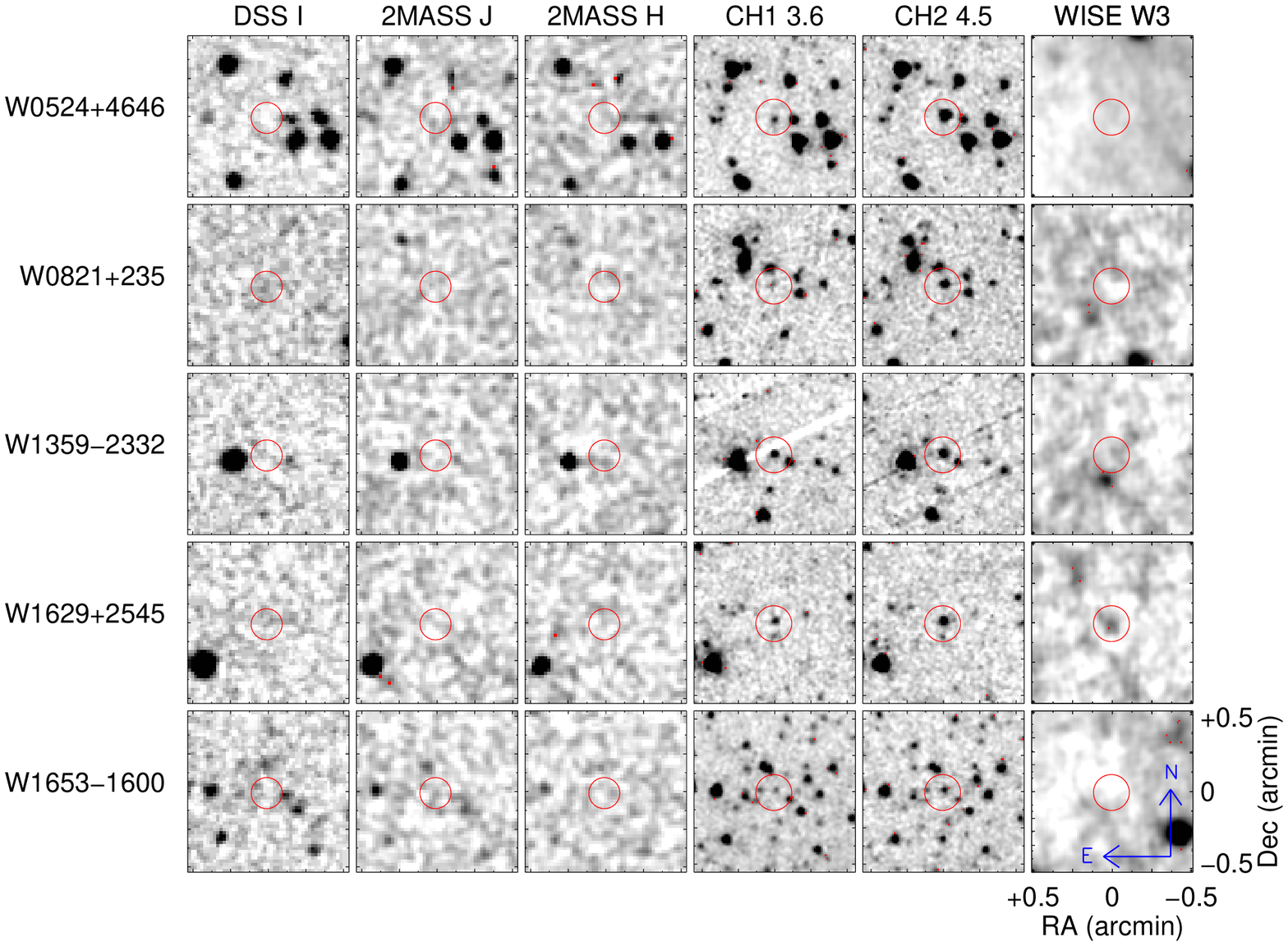}\\
\end{tabular}
\end{center}
\caption[CAPTION]{\label{fig:3} Figure 3 (continued)}
\end{figure*}

\begin{figure*}[!t]
\begin{center}
\begin{tabular}{l}
\includegraphics[scale=0.75,angle=90]{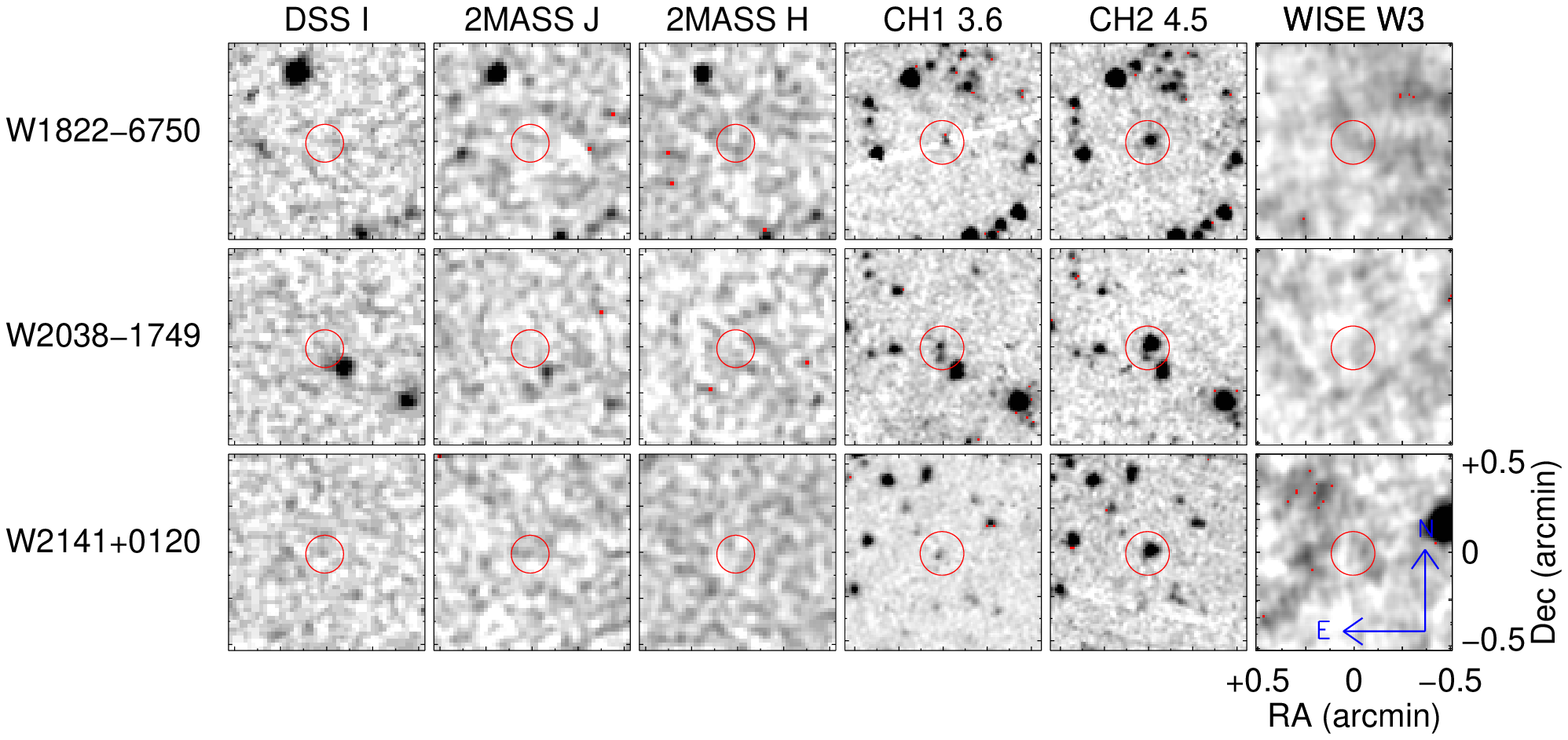}\\
\end{tabular}
\end{center}
\caption[CAPTION]{\label{fig:3} Figure 3 (continued)}
\end{figure*}

\clearpage


\end{document}